\documentclass[fleqn,usenatbib]{mnras}

\usepackage{newtxtext,newtxmath}
\usepackage[T1]{fontenc}

\DeclareRobustCommand{\VAN}[3]{#2}
\let\VANthebibliography\thebibliography
\def\thebibliography{\DeclareRobustCommand{\VAN}[3]{##3}\VANthebibliography}

\usepackage{graphicx}	% Including figure files
\usepackage{amsmath}	% Advanced maths commands
\usepackage{booktabs} 
\usepackage{caption}
\usepackage{subcaption}

\title[constraining terrestrial planets' formation]{The Moon-forming Impact as a Constraint for the Inner Solar System's Formation}

\author[Tong Fang et al.]{
Tong Fang$^{1}$\thanks{E-mail: tfang@shao.ac.cn},
Rongxi Bi$^{2}$,
Hui Zhang$^{1}$,
You Zhou$^{2}$,
Christian Reinhardt$^3$,
Hongping Deng$^{1}$
\\
$^{1}$Shanghai Astronomical Observatory, Chinese Academy of Sciences, 80 Nandan Road, Shanghai 200030, China\\
$^{2}$Research Center for Planetary Science, College of Earth and Planetary Sciences, Chengdu University of Technology, \\
No. 1, East Third Road, Erxianqiao, Chenghua District, Chengdu 610059, China\\
$^{3}$Department of Astrophysics, University of Zurich, Winterthurerstrasse 190, Zurich 8057, Switzerland
}

\date{Accepted XXX. Received YYY; in original form ZZZ}

\pubyear{\the\year{}}

\begin{document}
\label{firstpage}
\pagerange{\pageref{firstpage}--\pageref{lastpage}}
\maketitle

% Abstract of the paper
\begin{abstract}
The solar system planets are benchmarks for the planet formation theory. Yet two paradigms coexist for the four terrestrial planets: the prolonged collisional growth among planetesimals lasting $>100$ million years (Myr) and the fast formation via planetesimals accreting pebbles within 10 Myr. Despite their dramatic difference, we can hardly tell which theory is more relevant to the true history of the terrestrial planets' formation.
Here, we show that the Moon's origin puts stringent constraints on the pebble accretion scenario, rendering it less favourable. 
In the pebble accretion model, the one-off giant impact between proto-Earth and Theia rarely (probability $<$ 1\textperthousand) occurs at the right timing and configuration for the Moon formation. Even if a potential impact happens by chance, giant impact simulations reveal perfect mixing between proto-Earth and Theia, leaving no room for the observed primordial Earth mantle heterogeneity and the compositional difference, though small, between Earth and the Moon. Thus, the Earth-Moon system along other terrestrial planets should preferably form from chaotic collisional growth in the inner solar system. 

\end{abstract}

% Select between one and six entries from the list of approved keywords.
% Don't make up new ones.

%%%%%%%%%%%%%%%%%%%%%%%%%%%%%%%%%%%%%%%%%%%%%%%%%%

%%%%%%%%%%%%%%%%% BODY OF PAPER %%%%%%%%%%%%%%%%%%

\section{Introduction}

Sub-centimeter-sized dust, or pebbles, are ubiquitous in protoplanetary disks. They may concentrate into dense clumps via interaction with gas and collapse to form planetesimals beyond
tens of kilometers in size \citep{Nesvorny2019trans}. These planetesimals collide and merge to form protoplanets and then terrestrial planets \citep{Kokubo2000formation}.  This collisional growth has been the backbone of the classical formation theory of the solar system terrestrial planets, featuring a prolonged giant impact (collisions between protoplanets) stage \citep{Chambers2004planetary, Raymond2009building, Hansen2009formation}. Notably, the growth of Earth is concluded by a very energetic giant impact \citep{Quintana2016frequency}, which also leads to the formation of the Moon \citep{Canup2001origin, Deng2019primordial, Yuan2023moon}.  

Alternatively, planetesimals can accrete the swarm of tiny pebbles and grow efficiently, leading to a paradigm shift in planet formation theory \citep{Lambrechts2012, Johansen2017forming}. The newly proposed pebble accretion may supersede collisional growth in the terrestrial planets' formation, allowing Earth to form within several million years \citep{Johansen2021pebble}. However, we can not tell by which paradigm the terrestrial planets have formed given our limited understanding even of Earth's accretion time scale, varying from 3 Myr to 30 Myr \citep{Onyett2023silicon, Yin2002short, Yu2011fast,kleine2009hf}. Although some isotopic evidence suggests a limited contribution of outer solar system materials in terrestrial planets,  disfavoring the pebble accretion model \citep{burkhardt2021terrestrial}.

 Here, we show that the Moon-forming impact, which has been thoroughly explored in the classical planetesimal accretion model in terms of its configuration \citep{Jacobson2014highly, Quintana2016frequency}, results \citep{Canup2001origin} and implications \citep{Deng2019primordial, Yuan2023moon,zhou2024scaling}, can put stringent constraints on the pebble accretion scenario of terrestrial planets' formation, rendering it less favorable. We describe our numerical methods in Section \ref{sec:method}, where N-body simulations are employed to determine the statistics of potential Moon-forming impacts and giant impact simulations further reveal the consequences of these impacts. We present simulation results in Section \ref{sec:results} and conclude in Section \ref{sec:con}.

\section{Methods}
\label{sec:method}
\subsection{N-body simulation of planet formation}
\label{sec:nbody}

In the pebble accretion model, the terrestrial planets (Venus, Earth, Theia, Mars, but excluding Mercury) evolve along the same characteristic mass growth track determined by orbit migration and pebble accretion \citep{Johansen2021pebble}, where the mass of a protoplanet is related to its heliocentric distance $r$ by
\begin{equation}
\label{eq:growth}
    M(r)=M_{\text{max}}\left [1-\left ( \frac{r}{r_0}\right )^{1-\zeta} \right]^{3/4},
\end{equation}
where $r_0$ (1.6 Astronomical Unit, AU) is the starting point for inward migration, $M_{\text{max}}$ and $\zeta$  are parameters associated with the protoplanetary disk's properties, taken as 1.74 Earth Masses ($M_\oplus$) and 3/7. \citet{Johansen2021pebble} demonstrated that the four protoplanets could evolve to the right mass and position with almost exclusive pebble accretion. Mercury needs to form separately by accreting iron-rich pebbles \citep{Johansen2022nucleation} in contrast to planetesimal accretion models \citep[e.g.,][]{Fang2020extreme,2020ApJ...888L...1D}. However, they omitted the collision process between proto-Earth (0.6 $M_\oplus$)  and Theia (0.4 $M_\oplus$), simply citing that collision between sub-Earths may form the Moon \citep{Canup2012forming}.

To fill the gap, we simulated the evolution of the four protoplanets with the open-source N-body simulation code \emph{REBOUND} \citep{rebound} employing the \emph{Mercurius} integrator \citep{Rein2019hybrid} with an initial time step of 0.005 years. We adopted disk models of  \citet{Johansen2021pebble}
and evolved the system for 200 Myr to study the sub-Earth collision in terms of its likelihood, timing, and impact parameters. The gas surface density is 
\begin{equation}
    \Sigma_g= 610  (r/ \text{AU})^{-1/2} \text{exp}(-t/\tau) \quad \text{g/cm}^{2},
\end{equation}
where $\tau$ is set to be 1.5 Myr and the disk aspect ratio $h=0.024 (r/ \text{AU})^{2/7}$.  The type I migration of protoplanets is treated as an extra torque \citep{Cresswell2008three, D2010three} utilizing the \emph{REBOUNDx} extension \citep{Tamayo2020reboundx}. The planet migration rate in gaseous discs reads 
\begin{equation}
    \frac{dr}{dt}=-k_{\text{mig}}\frac{m}{M_\text{sun}}\frac{\Sigma_g r^2}{M_\text{sun}}h^{-2}v_K,
\end{equation}
where $m$ is the planet mass and $v_K$ is the Keplerian velocity at $r$. $k_{\text{mig}}$ is a constant related to the disk properties \citep{D2010three, Johansen2019planetary} taken as 3.708 here.
The eccentricity and inclination damping of protoplanets are included, similar to the migration \citep{Fang2023planetesimal}. 

We start from t=5 Myr when the disk is largely dispersed, and the four planets are nearly fully formed. Venus and Mars' masses grow along the mass growth track and are thus determined by their heliocentric distances according to equation \ref{eq:growth}. 
However, we turned off mass growth for the interacting proto-Earth and Theia due to their dynamic nature and the limited solids available when $t>5$ Myr. We follow the system until proto-Earth and Theia collide or up to 200 Myr, which is well beyond the expected Moon-forming event \citep{Halliday2023accretion}

\begin{figure}
    \includegraphics[width=\columnwidth]{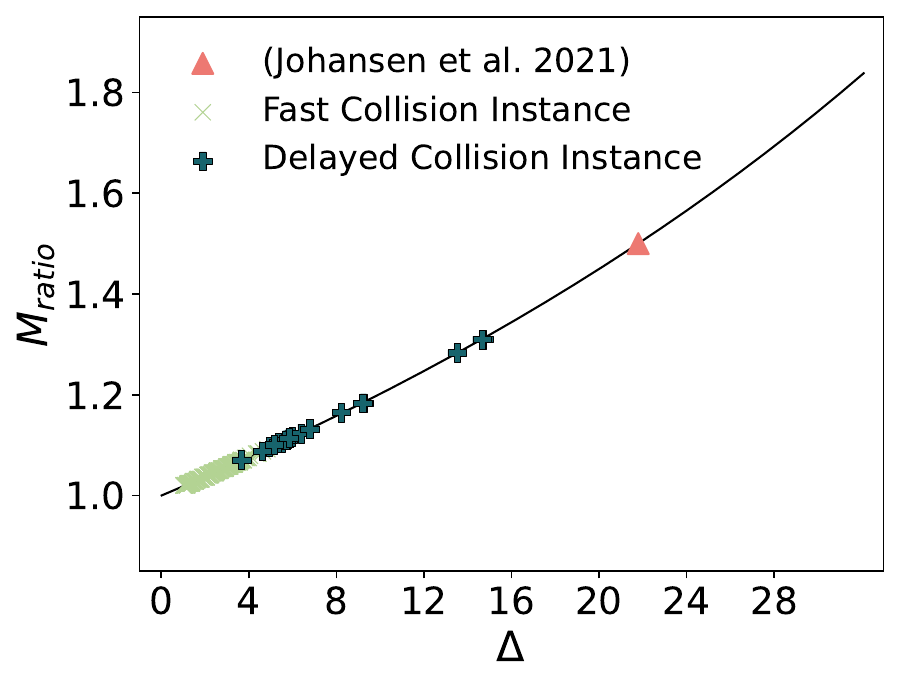}
    \caption{Proto-Earth to Theia mass ratio versus  $\Delta$ for simulations experienced proto-Earth and Theia collisions among the 2900 fiducial N-body simulations. Collisions within 1 Myr are regarded as fast collision instances and otherwise as delayed collision instances (see Fig. \ref{fig:Fig2}).} 
    \label{fig:Fig1}
\end{figure}

The evolution of proto-Earth and Theia is sensitive to their separation $\Delta$, measured by their 
mutual Hill radii 
\begin{equation}
  R_H=[(m_1+m_2)/3M_{\text{sun}}]^{1/3}[(a_1+a_2)/2], 
\end{equation}
where planet masses are $m_1, m_2$ and their semi-major axes are $a_1, a_2$. Notably, when  $\Delta<2\sqrt{3}$, an instability develops quickly \citep{Gladman1993}. With that in mind, we initialize pairs of proto-Earth and Theia with separations ($a_2-a_1$) ranging from 1 to 30 times $R_H$. In the meantime, $m_1+m_2=M_\oplus$ is required to form Earth at the proper mass after their fully merged collision. We neglected the minimal pebble accretion of the interacting proto-Earth and Theia in the largely dispersed disk from 5 to 200 Myr. These two conditions determine the starting mass and position of proto-Earth and Theia (Fig.~\ref{fig:Fig1}).  

\begin{figure*}
\centering
\includegraphics[scale=0.58]{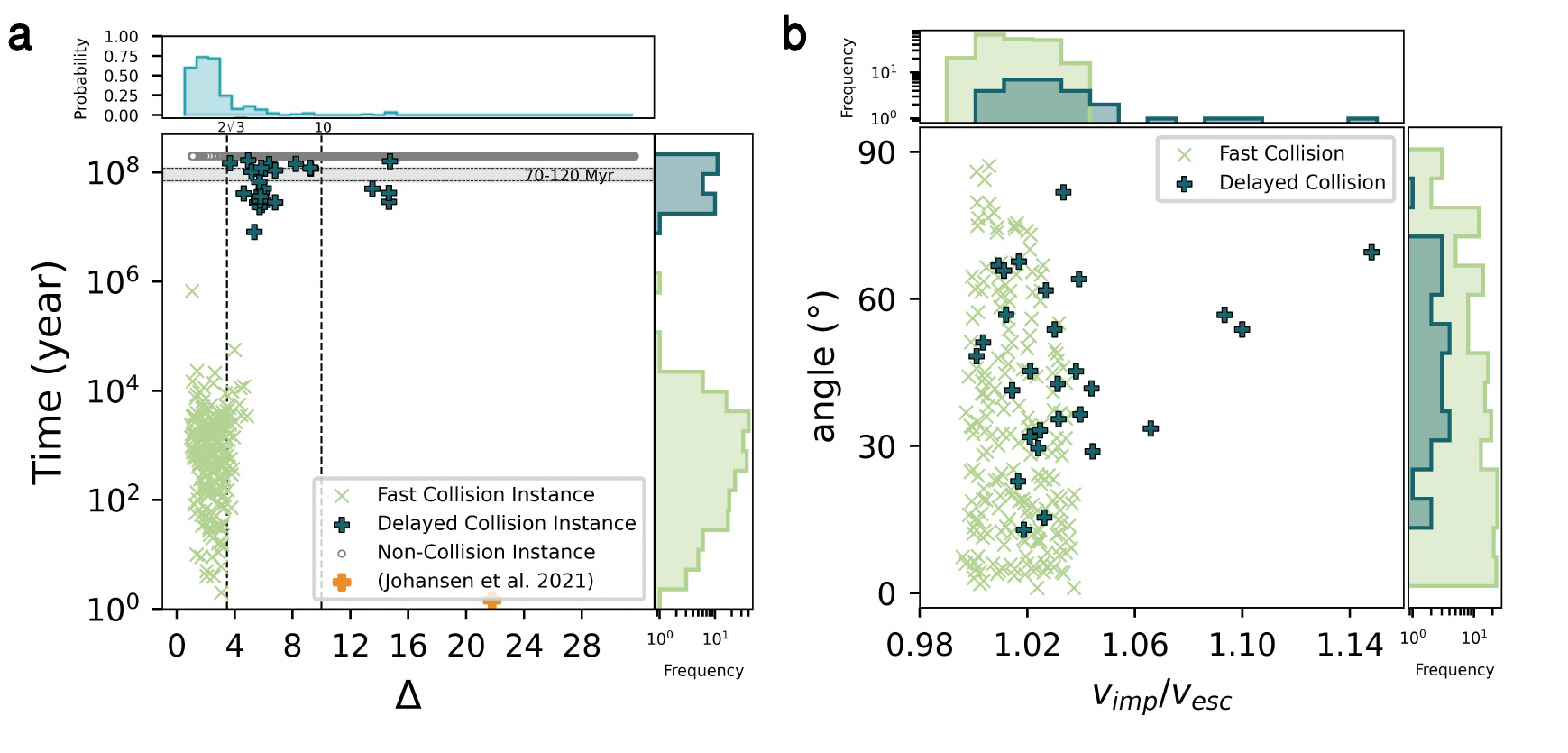}
    \captionsetup{textfont={stretch=1.}, singlelinecheck=false}
    \caption{The fate of proto-Earth and Theia in the 2900 fiducial protoplanet accretion simulations. \textbf{a}, For pairs of proto-Earth and Theia with initial separations $\Delta<2\sqrt{3}$, instability develops quickly, leading to collisions within 1 Myr for 67.5\% of the simulations (Fast Collision Instances).  Delayed collisions occur in only 1.5\% of the systems with initially $2\sqrt{3}<\Delta<20$,  and none for $\Delta>20$. Four instances occur at the right Moon formation timing.
    \textbf{b}, The collisions happen around the escape velocity between proto-Earth and Theia at angles almost uniformly distributed.}
    \label{fig:Fig2}
    
\end{figure*}
Our fiducial model consists of 2900 N-body simulations, 2300 covering $\Delta \in$ [1,21] uniformly and 600 covering $\Delta \in$ (21,31] uniformly for the proto-Earth and Theia pairs. Proto-Earth and Theia have initial eccentricities of 0.01 \citep{Johansen2021pebble}. Jupiter is also included in the simulation but with a fixed eccentricity of 0.05 and a fixed semi-major axis of 5 AU. Mercury is ignored here for its low mass and separated formation history \citep{Johansen2021pebble}. The argument of periapsis for all planets is assigned randomly. 

In addition, we conducted five groups of slightly different test simulations, each with 1200 runs uniformly covering $\Delta$ in the range [1,21]. These simulations tested the effects of eccentricity, Jupiter, and the gas disk on collision outcomes. The groups were as follows: 1. no initial eccentricity of proto-Earth and Theia, without (w/o) Jupiter, but with (w/) a dissipating gas disk; 2. no initial eccentricity, with Jupiter and a gas disk; 3. no initial eccentricity, with Jupiter, but without a gas disk; 4. initial eccentricity of 0.01, without Jupiter, but with a gas disk; 5. initial eccentricity of 0.01, with Jupiter, but without a gas disk.

We further tested the effects of migrating giant planets,  possibly due to an early giant planet instability \citep{clement2018mars,clement2019}, which can strongly excite the eccentricities of inner terrestrial planets and maximize the collision probability \citep{Agnor2012migration}. We incorporated the migrating giants into our models to study the orbital excitation of the four inner protoplanets and assess any increase in the collision rate between proto-Earth and Theia. We followed \cite{brasser2009} to simulate the orbital migration of Jupiter and Saturn, starting from an initial orbital period ratio $P_S$/$P_J$ = 1.5 with a migration e-folding timescale of 5 Myr. An eccentricity damping was applied to Saturn to mimic the effect of dynamical friction and avoid instability \citep{brasser2009,morbidelli2009}. Similarly to the standard scenario, we tested the following three setups: 1. no initial eccentricity of proto-Earth and Theia, without a dissipating gas disk; 2. no initial eccentricity, with a gas disk; 3. initial eccentricity of 0.01, with a gas disk. Each setup includes 4000 runs uniformly covering $\Delta$ in the range [2, 21].

We analyzed the probability of collisions and various parameters, including the time of giant impact, impact velocities, and impact angles. The final collision parameters also served as initial conditions for the giant impact simulations. 
    % We also compared the impact velocities to the mutual escape velocities, which indicated different theories and numerical models of the giant impact \cite{halliday2023}. The mutual escape velocity of the Earth and Theia is given by $v_{esc}=[2G(m_1+m_2)/(r_1+r_2)]^{1/2}$, where m means the mass and r means the radius. Moreover, we derived orbital parameters for Earth analogs formed from giant impacts in N-body simulations, facilitating comparison with the current orbital parameters of Earth.

%f resulting from the combined action of the magnetic field.}

\subsection{Giant impact simulations}
\label{sec:impact}
The giant impacts between the sub-Earths are simulated with the GIZMO code employing the Meshless Finite Mass (MFM)  method \citep{Hopkins2015}, which is a Godunov-type Lagrangian method free of artificial viscosity needed in SPH for capturing shocks. Thus, MFM captures the turbulence and associated mixing during the impacts more faithfully than SPH at comparable resolutions \citep{Deng2019enhanced}. We use the public Python package \emph{SEAGEN} to build sub-Earth models \citep{Kegerreis2019planetary}, employing an updated version of the ANEOS equation of state \citep{Thompson1974improvements, Stewart2020shock} for mantle (forsterite, 70wt\%) and core (iron, 30wt\%). The code has been applied to similar Moon formation simulations involving a Mars-sized impactor, i.e., the canonical Moon-forming scenario, revealing a layered post-impact proto-Earth \citep{Deng2019primordial} in line with SPH simulations utilizing 100 times more resolution elements \citep{Yuan2023moon}. Here, each simulation is resolved by one/four million fluid elements, which is adequate for MFM to characterize the post-impact proto-Earth \citep{Deng2019primordial}.

We followed the impacts for over 35 hours until the protolunar disks were well established. We identified the protolunar disks by iterative solving for gravitationally bound debris with periapsis larger than the post-impact planet radius \citep{Canup2001scaling}. The mass of the Moon expected to form within the protolunar disk is estimated based on an empirical formula obtained by N-body simulations following \citet{Canup2012forming}, although the vapor-dominated disk may not form the Moon at all \citep{Nakajima2022large, Nakajima2024limited}.

\section{Results}
\label{sec:results}
\subsection{Impact Statistics}
Four protoplanets (excluding Mercury) are formed along the same growth track in the pebble accretion scenario. The Moon must result from the one-off collision between proto-Earth and Theia (a hypothetical protoplanet that collided with the proto-Earth to form the Moon) \citep{Johansen2021pebble}. However, to form the Earth-Moon system, several key constraints must be satisfied: 1) the masses of proto-Earth and Theia add up to about one Earth mass, 2) the two protoplanets are close enough so their orbits become unstable \citep{Gladman1993}, 3) the giant impact occurs at the right timing for the Moon formation (70-120 Myr) \citep{Halliday2023accretion}. We carried out planetary accretion simulations to reveal the characteristics of the Moon-forming impact in the pebble accretion model for terrestrial planets' formation (see Section \ref{sec:nbody}) and summarised the results in Fig. \ref{fig:Fig2}. 

When the orbital separation between proto-Earth and Theia ($\Delta$, measured by their mutual Hill radii, see Section \ref{sec:nbody}) is smaller than $2\sqrt{3}$, the two half-Earths become unstable \citep{Gladman1993} and likely collide within 1 Myr. However, the collision probability declines to 1.5\% for systems with $2\sqrt{3}< \Delta<20$ (see Fig.~\ref{fig:Fig2}a). The close separation implies a similar accretion history and translates to similar masses in the two protoplanets \citep{Johansen2021pebble}. In fact, sub-Earth impacts with a mass ratio of 1.5, as cited in \citet{Johansen2021pebble}, never happened in our long-term simulations (see Fig. \ref{fig:Fig1}). On the other hand, the Moon is believed to form between 70-\textbf{120} Myr after Ca-Al-rich inclusions (CAIs)  formation \citep{Halliday2023accretion}, and only four instances of our 2900 simulations occur at the right timing. 

\begin{figure}
    \centering
    \begin{subfigure}[b]{\columnwidth}
        \centering
        \includegraphics[width=\linewidth]{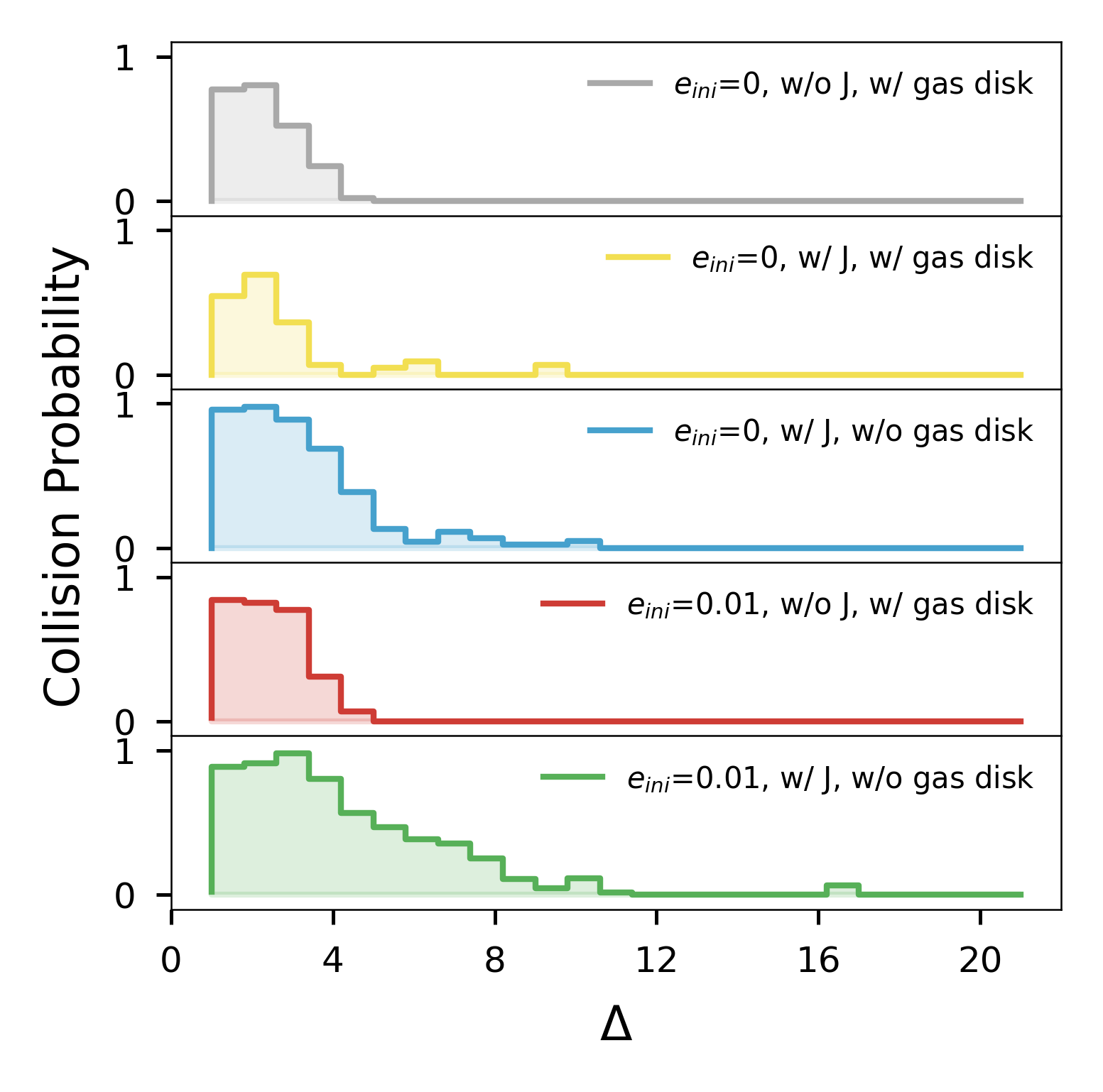}
        \put(-230, 230){\large \textbf{a}}
    \end{subfigure}
    \vspace{-0.5cm}
    \begin{subfigure}[b]{\columnwidth}
        \centering
        \includegraphics[width=\columnwidth]{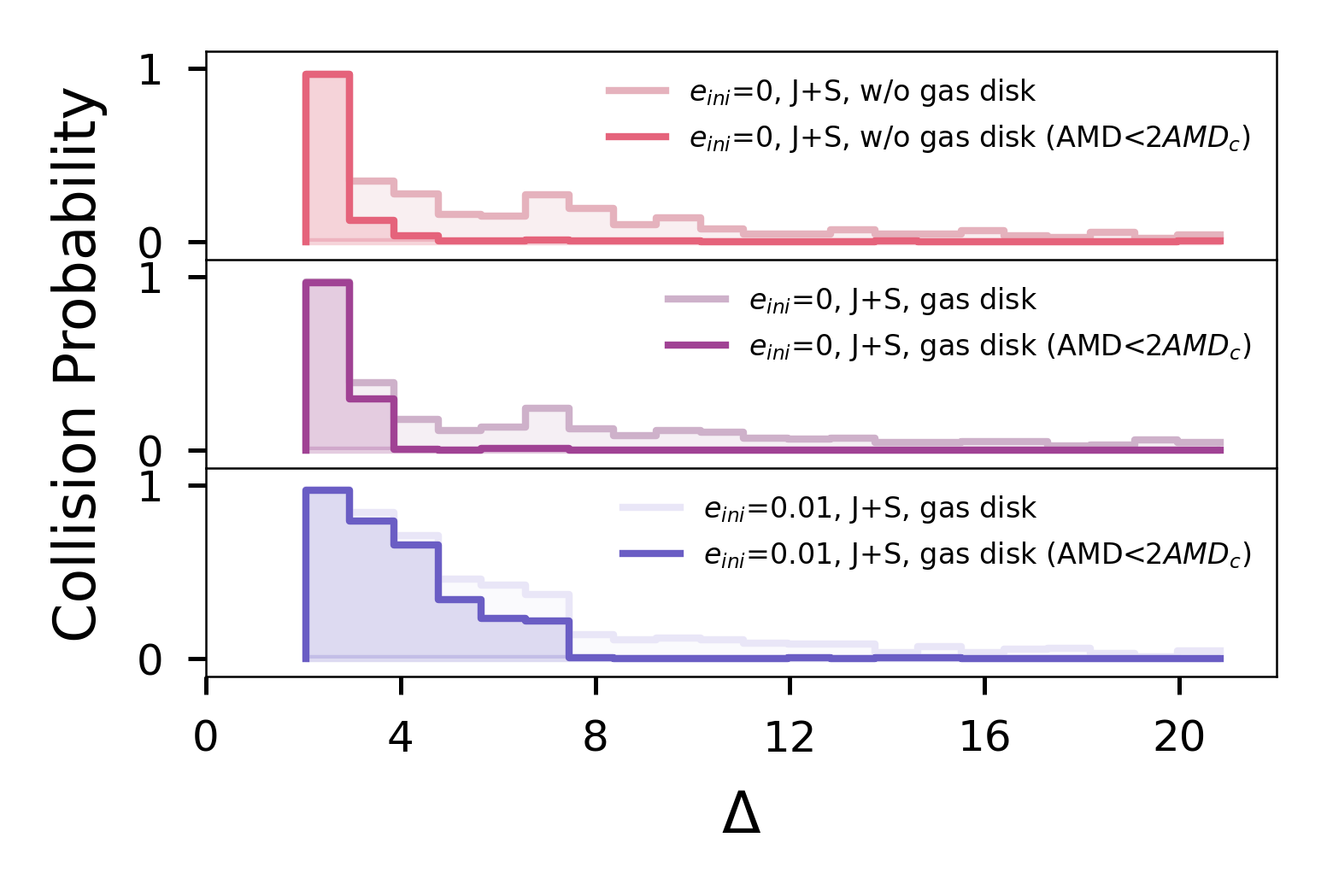}
        \put(-230, 160){\large \textbf{b}}
    \end{subfigure}
    \caption{The collision probability between proto-Earth and Theia for different groups of test simulations (see Section \ref{sec:nbody}), to be compared with the fiducial simulations in Fig. \ref{fig:Fig2}a. \textbf{a}, Five groups of simulations testing the effects of Jupiter, gas disk, and initial eccentricity, with each group consisting of 1200 simulations covering $\Delta \in$ [1,21] uniformly. \textbf{b}, Simulations with migrating giant planets, with each group consisting of 4000 simulations covering $\Delta \in$ [2,21] uniformly. The light-colored areas represent collision cases without AMD constraints, and the dark-colored areas represent cases with AMD limited to less than twice the current AMD.}
    \label{fig:Fig3}
\end{figure}

\begin{table*}
	\centering
	\caption{Giant impact simulations. Table columns: simulation number (ID), with the subscript `s' denoting successful Moon-forming impact that may form an iron-poor Moon close to lunar mass \citep{Canup2012forming} ($0.8 < M_{F,L}/M_L < 2.0$) and the subscript `h' indicates the high-resolution simulations; the mass ratio between proto-Earth and Theia ($M_{pE}/M_T$) and their separation ($\Delta$) on the mass growth track; the impact parameter ($b$), i.e., the \emph{sine} of the impact angle; impact velocity relative to the escape velocity ($v_{\text{imp}}/v_{\text{esc}}$); disk mass in lunar mass ($M_d/M_L$); iron mass in the disk relative to lunar mass ($M_{\text{d,iron}}/M_L$); angular momentum of the disk relative to the angular momentum of the Earth-Moon system ($L_d/L_{EM}$); final angular momentum of the planet and disk relative to the angular momentum of the Earth-Moon system ($L_F/L_{EM}$); mass fraction of Theia-origin material in the protolunar disk ($M_{To}/M_d$); predicted mass of the Moon relative to lunar mass ($M_{F,L}/M_L$).}
	\label{tab_fwsc}
	\begin{tabular*}{0.9\linewidth}{@{}lllllllllll@{}} % 11 columns, alignment for each
    \toprule
        ID &$M_{pE}/M_T$ &$\Delta$& $b$ & $v_{imp}/v_{esc}$ & $M_d/M_L$ & $M_{d,iron}/M_L$ & $L_d/L_{EM}$ & $L_F/L_{EM}$ & $M_{To}/M_d$ & $M_{F,L}/M_L$   \\ \hline
    \midrule 
        $1$ & 1.00  & 0.00  & 0.35  & 1.00  & 0.00  & 0.00  & 0.00  & 1.34  & 1.00  & 0.00   \\ 
        $2$ & 1.00  & 0.00  & 0.45  & 1.00  & 0.82  & 0.00  & 0.11  & 1.64  & 0.57  & 0.12   \\ 
        $3_{s}$ & 1.00  & 0.00  & 0.55  & 1.00  & 3.04  & 0.02  & 0.44  & 1.82  & 0.52  & 0.87   \\ 
        $4_{s,h}$ & 1.00  & 0.00  & 0.55  & 1.00  & 2.47  & 0.08  & 0.38  & 1.93  & 0.51  & 0.89   \\ 
        $5_{s}$ & 1.00  & 0.00  & 0.60  & 1.00  & 3.19  & 0.04  & 0.49  & 2.23  & 0.49  & 1.15   \\ 
        $6_{s}$ & 1.00  & 0.00  & 0.65  & 1.00  & 3.63  & 0.07  & 0.56  & 2.13  & 0.51  & 1.38   \\ 
        $7_{s,h}$ & 1.00  & 0.00  & 0.65  & 1.00  & 2.09  & 0.04  & 0.34  & 2.28  & 0.50  & 0.97   \\ 
        $8_{h}$ & 1.00  & 0.00  & 0.65  & 1.02  & 5.43  & 0.23  & 0.86  & 2.21  & 0.46  & 2.41   \\ 
        $9_{s}$ & 1.00  & 0.00  & 0.70  & 1.00  & 2.38  & 0.02  & 0.42  & 2.39  & 0.50  & 1.37   \\ 
        $10_{s,h}$ & 1.00  & 0.00  & 0.70  & 1.00  & 2.68  & 0.05  & 0.45  & 2.42  & 0.50  & 1.31   \\ 
        $11_{h}$ & 1.00  & 0.00  & 0.70  & 1.02  & 7.11  & 0.08  & 1.21  & 2.48  & 0.47  & 4.12   \\ 
        $12_{h}$ & 1.00  & 0.00  & 0.70  & 1.04  & 5.24  & 0.03  & 0.87  & 2.39  & 0.50  & 2.75   \\ 
        $13_{s}$ & 1.00  & 0.00  & 0.80  & 1.00  & 4.19  & 0.01  & 0.68  & 2.40  & 0.50  & 1.98   \\ 
        $14_{s,h}$ & 1.12  & 6.18  & 0.65  & 1.00  & 1.97  & 0.03  & 0.35  & 2.35  & 0.48  & 1.10   \\ 
        $15_{s,h}$ & 1.12  & 6.18  & 0.70  & 1.00  & 2.36  & 0.03  & 0.39  & 2.45  & 0.50  & 1.10   \\ 
        $16_{s}$ & 1.17  & 8.55& 0.65  & 1.02  & 3.93  & 0.10  & 0.65  & 2.23  & 0.53  & 1.88   \\ 
        $17_{s}$ & 1.17  & 8.55& 0.65  & 1.04  & 2.79  & 0.01  & 0.47  & 2.37  & 0.55  & 1.38   \\ 
        $18$ & 1.17  & 8.55& 0.65  & 1.06  & 5.36  & 0.07  & 0.89  & 2.31  & 0.50  & 2.78   \\ 
        $19_{h}$ & 1.18  & 9.01  & 0.65  & 1.00  & 4.33  & 0.11  & 0.71  & 2.36  & 0.56  & 2.02   \\ 
        $20_{s,h}$ & 1.18  & 9.01  & 0.70  & 1.00  & 2.37  & 0.04  & 0.41  & 2.50  & 0.49  & 1.29   \\ 
        $21$ & 1.18  & 9.01  & 0.74  & 1.00  & 6.39  & 0.09  & 1.12  & 2.67  & 0.42  & 3.84   \\ 
        $22$ & 1.20  & 9.92  & 0.60  & 1.02  & 0.90  & 0.01  & 0.15  & 2.18  & 0.48  & 0.41   \\ 
        $23_{s}$ & 1.20  & 9.92  & 0.60  & 1.04  & 3.23  & 0.03  & 0.49  & 2.07  & 0.53  & 1.11   \\ 
        $24$& 1.20  & 9.92  & 0.60  & 1.06  & 3.53  & 0.11  & 0.59  & 2.07  & 0.58  & 1.79   \\ 
        $25_{h}$ & 1.20  & 9.92  & 0.75  & 1.00  & 4.78  & 0.02  & 0.86  & 2.50  & 0.60  & 3.11   \\ 
        $26$ & 1.27  & 12.98  & 0.70  & 1.00  & 1.10  & 0.00  & 0.21  & 2.36  & 0.49  & 0.73   \\ 
        $27$ & 1.27  & 12.98  & 0.83  & 1.00  & 3.95  & 0.03  & 0.66  & 2.27  & 0.36  & 2.05   \\ 
        $28_{s,h}$ & 1.31  & 14.64  & 0.70  & 1.00  & 1.99  & 0.01  & 0.35  & 2.38  & 0.49  & 1.11   \\ 
        $29$ & 1.31  & 14.64  & 0.83  & 1.00  & 5.78  & 0.07  & 1.15  & 2.86  & 0.74  & 4.90   \\ 
    \bottomrule
	\end{tabular*}
\end{table*}

The impact angle is almost uniformly distributed, while the impact velocity is close to their mutual escape velocity (Fig.~\ref{fig:Fig2}b). The delayed impacts show a relatively high impact velocity due to the complete dispersal of the gas disk and the absence of eccentricity damping. 

We tested the findings against different protoplanetary disk properties (see Section \ref{sec:nbody}) and observed similar results (Fig.~\ref{fig:Fig3}) to the fiducial models (Fig. \ref{fig:Fig2}a). Widely separated proto-Earth (0.6 $M_\oplus$)  and Theia (0.4 $M_\oplus$) pairs, as envisaged by \cite{Johansen2021pebble}, never collided in our 2900 N-body simulations.

In Fig.\ref{fig:Fig3}a, the presence of Jupiter is essential for wide separation collisions with $\Delta>10$. The highest collision occurrence rate is achieved when eccentric proto-Earth and Theia evolve in gas-free environments. However, the collision probability for proto-Earth and Theia pairs with $\Delta>10$ is always minimal (Fig.~\ref{fig:Fig3}a). With migrating giants, the inner terrestrial planets are over-excited in most simulations. Only simulations with angular momentum deficit (AMD) less than twice the current AMD value (excluding Mercury)  can be regarded as successful \citep[e.g.,][]{clement2018mars}. In Fig. \ref{fig:Fig3}b, collisions between proto-Earth and Theia occur easily when $\Delta$ is less than $2\sqrt{3}$. However, as $\Delta$ increases, the probability of Earth-Theia collisions significantly decreases, especially when applying the aforementioned AMD criterion. There are very few Earth-Theia collisions with $\Delta$ exceeding 8 in systems with an appropriate degree of orbital excitation. 

In the pebble accretion model for terrestrial planets' formation, the Moon-forming impact only occurs for close proto-Earth and Theia pairs featuring similar masses, and the late collisions (70-120 Myr) only occur when the separation is between $2\sqrt{3}$ and 16 Hill radii. The probability that such a collision happens for separation in this range is less than 3\textperthousand  ~  (only 1/3 of them can potentially form the Moon, see section \ref{sec:result2}). If the time window is extended to 40-200 Myr, the probability of late impact remains below 1.2\%.

\subsection{Impact Simulations}
\label{sec:result2}
To reveal the consequences of the above impacts, we performed a series of hydrodynamic simulations (see Section \ref{sec:impact}) for the half-Earth impacts (Table~\ref{tab_fwsc}) covering the impact parameter space identified above. Our models are strictly confined to half-Earths with nearly identical mass and low-velocity impacts, different from the general sub-Earth impact model of Moon formation \citep{Canup2012forming, Timpe2023systematic, meier2024systematic}. Here, we focus on the state of the post-impact proto-Earth, the protolunar disk, and their geochemical implications \citep{Deng2019primordial}.

\begin{figure*}
\centering
\includegraphics[scale=0.5]{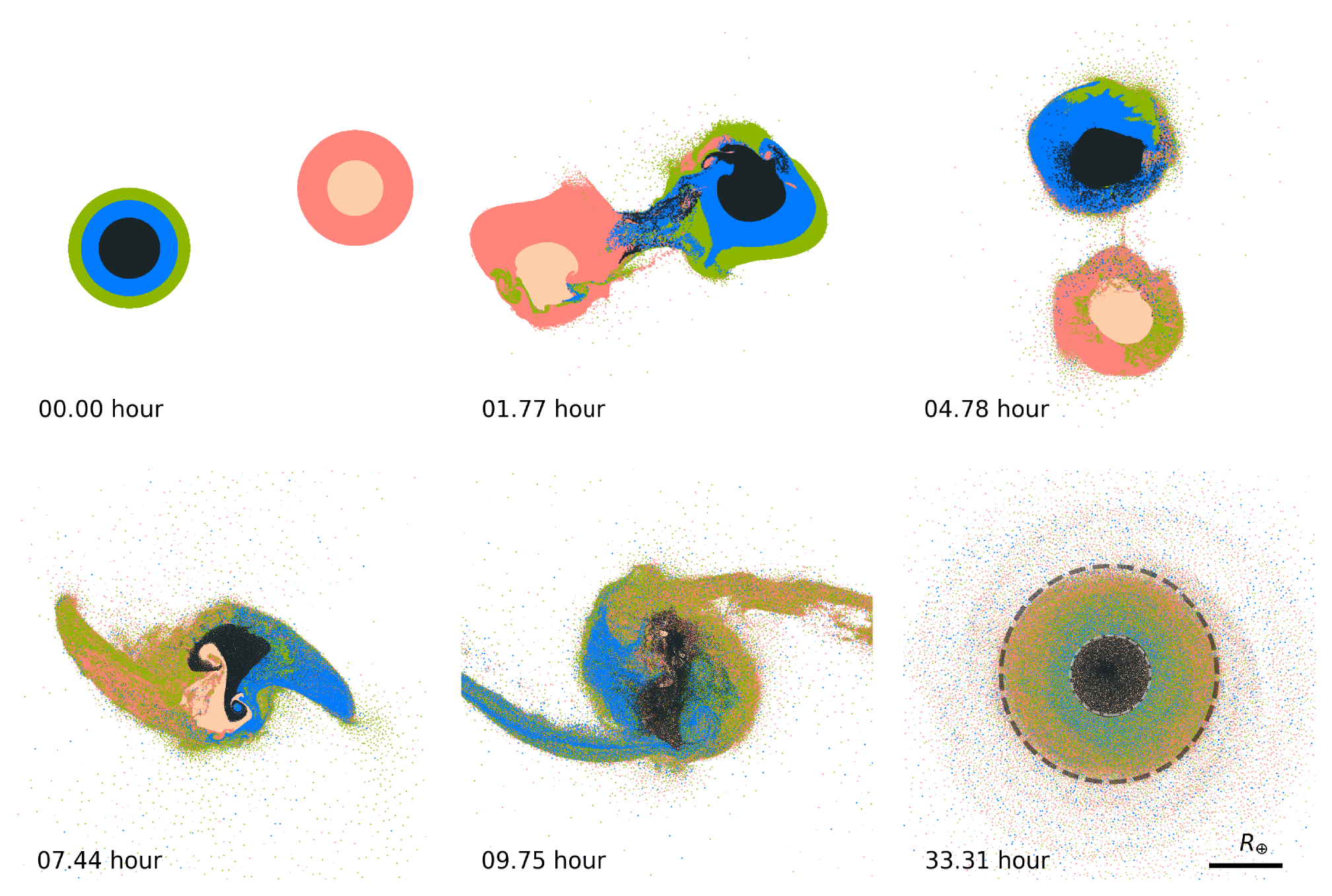}
    \captionsetup{textfont={stretch=1.0}, singlelinecheck=false}
    \caption{Simulation of a half-Earth impact (run 20 of Table~\ref{tab_fwsc}). The slice plots show only materials near the equatorial plane with a thickness of 0.2 present-day Earth Radius ($R_\oplus$). 
    The colors trace the mixing of materials between the target (proto-Earth) and impactor (Theia), where different colors indicate the core and mantle materials, and the lower and upper half of the target's mantle are further distinguished by blue and green. The dashed lines denote the core-mantle boundary and the surface of the post-impact target reaching 1.6 $R_\oplus$ due to its fast spin. 
    A supplementary movie can be downloaded at \url{https://drive.google.com/file/d/1DdH59kOv3Y3pc5PQXQnSIInDGfO3jZ64/view?usp=drive_link}.}
    \label{fig:Fig4}
    
\end{figure*}

The models that produce the right lunar mass (see Section \ref{sec:impact} and Table~\ref{tab_fwsc}) typically have an initial angular momentum 2 to 2.7 times that of the current Earth-Moon system ($L_{EM}$) \citep{Timpe2023systematic}. An impact angle between 30 and 45 degrees is desirable, accounting for 1/3 of the delayed collisions in Fig.~\ref{fig:Fig2}b. When the target (proto-Earth) and impactor (Theia) have identical masses, they are indistinguishable and expected to contribute equally to the post-impact structure everywhere. Efficient mixing also happens for simulations with noticeable mass differences. In Fig.~\ref{fig:Fig4}, we trace the material distribution in an impact with a target-to-impactor mass ratio of 1.18 ($\Delta=9.01$).  The impactor disrupts the target upon first contact and exposes the lower half of the target mantle. They then merge, and Kelvin-Helmholtz type of instability leads to efficient mixing of materials of different origins. After the impact, the oblate fast-spinning Earth has an equatorial radius of 1.6 times the present-day Earth radius ($R_\oplus$) and an angular momentum of more than 2$L_{EM}$. Yet it is controversial whether this excessive angular momentum can be removed \citep{Rufu2020tidal}. 
%Such impacts face two additional difficulties: it is controversial whether this excessive angular momentum can be removed\cite{Rufu2020tidal}, and the resulting vapor-rich disks might not produce massive satellites due to strong gas drag on the moonlets\cite{Nakajima2024limited}. 

\begin{figure}
  \includegraphics[width=\columnwidth]{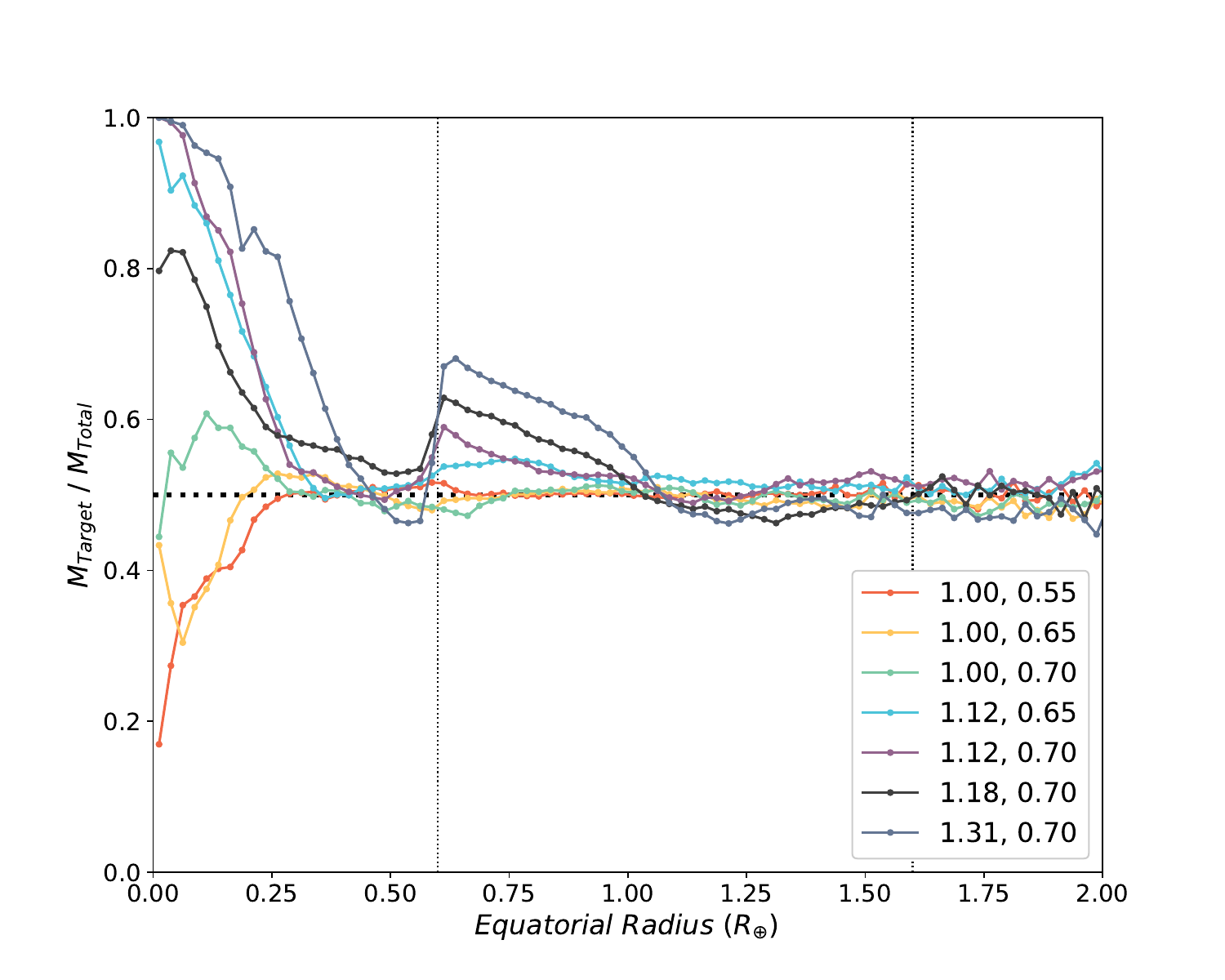}
    \caption{The fraction of target-origin material in the post-impact structure at different radii in successful high-resolution simulations (see Table~\ref{tab_fwsc}). The labels indicate the target-to-impactor mass ratio and the impact parameter. The vertical dotted lines indicate the core-mantle boundary and the planet's surface (Fig.~\ref{fig:Fig4}), while the horizontal dotted line provides a reference value of 0.5.}
    \label{fig:Fig5}
\end{figure}

\begin{figure}
 \includegraphics[width=\columnwidth]{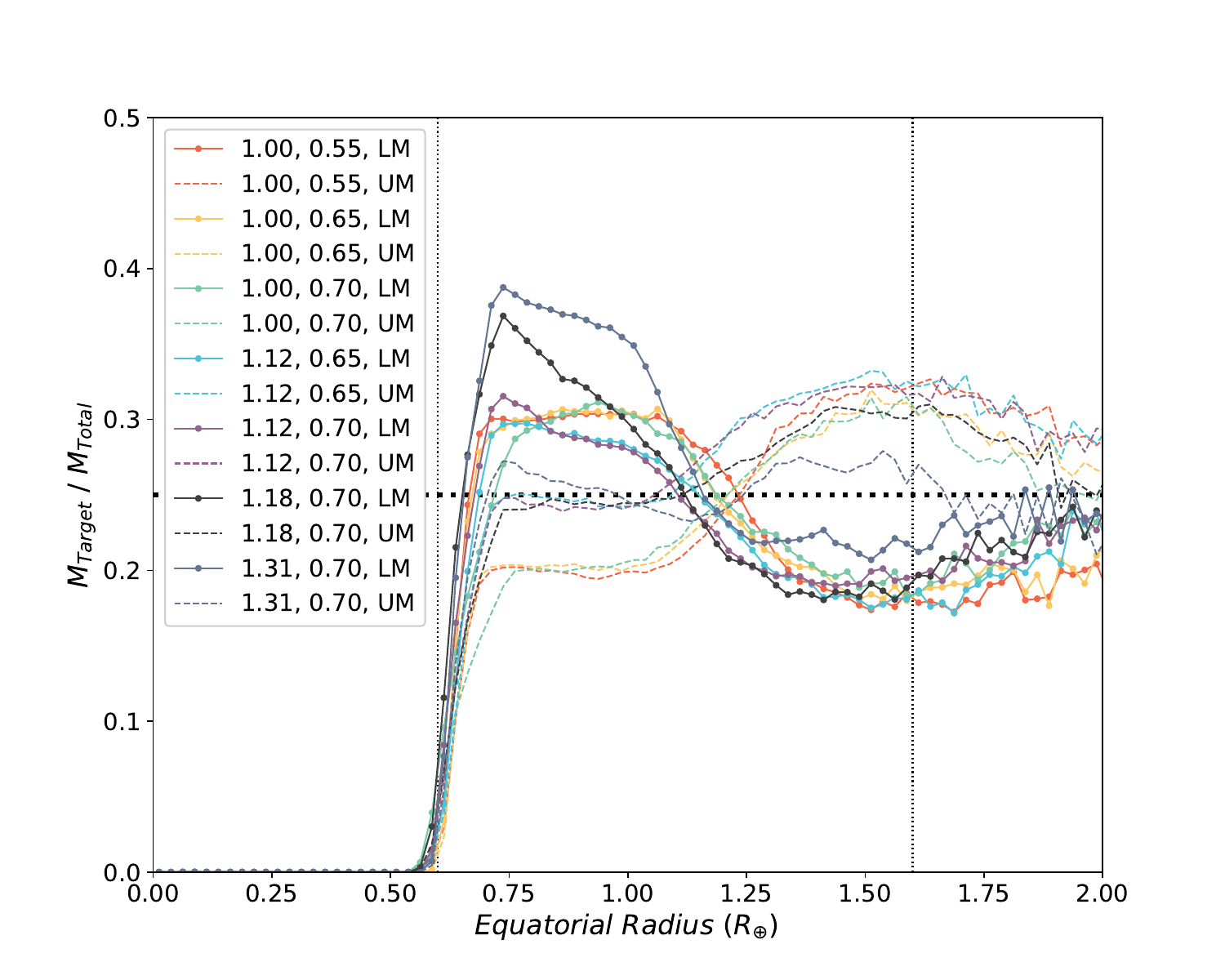}
  \caption{The redistribution of the target's lower mantle (LM) and upper mantle (UM) materials after the impact for successful high-resolution simulations in Table~\ref{tab_fwsc}. The labels indicate the target-to-impactor mass ratios and the impact parameters. The vertical dotted lines indicate the core-mantle boundary and the planet's surface (Fig.~\ref{fig:Fig4}), while the horizontal dotted line provides a reference value of 0.25. }
  \label{fig:Fig6}
\end{figure}

We calculated the compositional profile of the post-impact target to quantify the mixing state. The oblate spheroid has a major axis of about 1.25 times its minor axis. As a result, we calculate compositional and thermal profiles by averaging quantities within concentric ellipsoids resembling the planet's shape. We use the major axis (equatorial radius) as a reference for the radius in later plots.
The target slightly dominates the basal mantle in simulations with a noticeable target-impactor mass imbalance (Fig.~\ref{fig:Fig5}). However, the impactor and target are well mixed beyond one Earth radius extending to the protolunar disk. Remarkably, the lower mantle material of proto-Earth can contribute significantly to the protolunar disk, matching the contribution of the upper mantle material (Fig.~\ref{fig:Fig6}). The lower and upper mantle exchange is also prominent since the impact turns proto-Earth inside-out at some stages (Fig.~\ref{fig:Fig4}).  As a result, if any primordial reservoirs predate the Moon-forming impact, they would have been efficiently mixed during the impact and further homogenized by convection in the fully molten post-impact early Earth \citep{Nakajima2015melting}. This thorough mixing %is in tension
is difficult to reconcile
with the observed primordial heterogeneities in Earth that predate the Moon-forming impact \citep{Mukhopadhyay2012early, Touboul2012182w, Rizo2016preservation}. Later on, heterogeneities within Earth's deep mantle may also arise from early differentiation processes \citep{labrosse2007crystallizing,lee2010upside}, the subduction of oceanic crust \citep{christensen1994segregation,yuan2024giant} and magnesium oxide exsolution from the core \citep{Deng2023primordial,rizo2019182w}.  However, the survival of primordial  mantle heterogeneity  provides a compelling explanation for the size of Earth's basal mantle anomalies \citep{Yuan2023moon}, which is absent in other mechanisms.

The degree of mixing in the protolunar disk is also extremely high. Successful simulations predict a fraction of Theia-origin material between 48-55\% in the disk. This perfect mixing and similar accretion history of proto-Earth and Theia leave little room for any difference in composition between Earth and the Moon. High-resolution Oxygen isotope measurements also reveal a compositional difference between Earth and the Moon  of 22 ppm in terms of $\Delta ^{17}\text{O}$ \citep{Cano2020distinct}, where $\Delta^{17}\text{O}=(\delta^{17}\text{O}-\lambda\cdot\delta^{18}\text{O})\times10^3$ \citep[but see][]{young2016oxygen}. Following the isotope mass balance calculation of \citet{Cano2020distinct} (see their Extended Data Fig. 4), we can estimate the initial compositional difference in proto-Earth and Theia given any impact mixing model. Assuming the thorough mixing in run 20, proto-Earth and Theia would have an initial isotope difference of 695 ppm, which is even larger than the Earth-Mars Oxygen composition difference. While in the pebble accretion model, proto-Earth and Theia are the closest siblings, likely possessing a smaller difference in Oxygen composition than the Earth-aubrites difference (22 ppm) \citep{Greenwood2018oxygen}. Nevertheless, more lunar samples across different regions are needed to reveal the intrinsic compositional difference between Earth and the Moon, avoiding bias due to compositional heterogeneity, for example, caused by the late veneer.

\section{Conclusions}
\label{sec:con}
We have combined N-body and giant impact simulations to study the Moon-forming impact in the pebble accretion model for the inner solar system's formation. In this model, we find Moon-forming impacts involving very similar proto-Earth and Theia on close-by orbits. Let alone the uncertainty in how they get there, the chance for a proper Moon-forming impact at the right timing and angle is below 1\textperthousand. With thorough mixing, such half-Earth impacts destroy primordial heterogeneities and lead to identical (not similar) isotopic compositions on Earth and the Moon, which is in tension with recent geochemical observations. We thus conclude the pebble accretion model can hardly explain the Earth-Moon system. The Moon, as an indispensable member of the inner solar system, helps differentiate models of the terrestrial planets' formation. 

\section*{Acknowledgements}

We would like to thank the anonymous reviewers for the valuable comments and suggestions, which have greatly enhanced this manuscript. H.D. and H.Z. acknowledge support from the Chinese Academy of Science via a talent program. Simulations were performed at the National Supercomputing Center in Jinan, and we are grateful for the help of the support staff. T.F. and H.Z. are further supported by the National Science Foundation of China  for Young Scientists (NSFC grant No.12403073) and the National Science Foundation of China (NSFC grant No.12073010, 11933001). Y.Z. is supported by the Sichuan Provincial Natural Science Foundation (grant No. 2023NSFSC0278) and the National Science Foundation of China (NSFC grant No. 4224114).

%%%%%%%%%%%%%%%%%%%%%%%%%%%%%%%%%%%%%%%%%%%%%%%%%%
\section*{Code Availability}
The rebound code for N-body simulations is available at \url{https://rebound.readthedocs.io/en/latest/} with its extensions, including planet orbital migration, available at \url{https://reboundx.readthedocs.io/en/latest/index.html}. The hydro-code GIZMO, including both SPH and MFM algorithms, is made available by its author, Philip Hopkins, at \url{http://www.tapir.caltech.edu/~phopkins/Site/GIZMO.html}

\section*{Data Availability}
The data files that support our analysis will be made available upon reasonable request to the corresponding author.

%%%%%%%%%%%%%%%%%%%% REFERENCES %%%%%%%%%%%%%%%%%%

% The best way to enter references is to use BibTeX:

\bibliographystyle{mnras}
\bibliography{pebblemoon} % if your bibtex file is called example.bib

% Alternatively you could enter them by hand, like this:
% This method is tedious and prone to error if you have lots of references
%\begin{thebibliography}{99}
%\bibitem[\protect\citeauthoryear{Author}{2012}]{Author2012}
%Author A.~N., 2013, Journal of Improbable Astronomy, 1, 1
%\bibitem[\protect\citeauthoryear{Others}{2013}]{Others2013}
%Others S., 2012, Journal of Interesting Stuff, 17, 198
%\end{thebibliography}

%%%%%%%%%%%%%%%%%%%%%%%%%%%%%%%%%%%%%%%%%%%%%%%%%%

%%%%%%%%%%%%%%%%% APPENDICES %%%%%%%%%%%%%%%%%%%%%

%%%%%%%%%%%%%%%%%%%%%%%%%%%%%%%%%%%%%%%%%%%%%%%%%%

% Don't change these lines
\bsp	% typesetting comment
\label{lastpage}
\end{document}